\documentclass[fleqn,usenatbib]{mnras}

\usepackage{newtxtext,newtxmath}
\usepackage{comment}

\usepackage[T1]{fontenc}

\DeclareRobustCommand{\VAN}[3]{#2}
\let\VANthebibliography\thebibliography
\def\thebibliography{\DeclareRobustCommand{\VAN}[3]{##3}\VANthebibliography}

\usepackage{graphicx}	
\usepackage{amsmath}	
\usepackage{xcolor}

\title[Pop~III XRBs in the local Universe]{Detectability of Population~III stellar remnants as X-ray binaries from tidal captures in the local Universe}

\author[R. Husain et al.]{
Rabia Husain,$^{1, 2}$\thanks{E-mail: r.husain100@gmail.com}
Boyuan Liu$^{1}$
and Volker Bromm$^{1}$
\\
$^{1}$Department of Astronomy, University of Texas, TX 78712, USA\\
$^{2}$Jefferson Laboratory of Physics, Harvard University, Cambridge, MA 02138
}

\date{Accepted XXX. Received YYY; in original form ZZZ}

\pubyear{2021}

\begin{document}
\label{firstpage}
\pagerange{\pageref{firstpage}--\pageref{lastpage}}
\maketitle

\begin{abstract}
{
We assess the feasibility of detecting the compact object remnants from Population~III (Pop~III) stars in nearby dense star clusters, where they become 
luminous again as X-ray binaries (XRBs) and tidal disruption events (TDEs) via strong tidal encounters.} Analytically modelling the formation of Pop~III stars, coupled with a top-heavy initial mass function predicted by numerical simulations, we 
{
derive the number of (active) Pop~III XRBs and TDEs in the present-day Milky Way (MW) nuclear star cluster as $\sim 0.06-0.3\ $ and $\lesssim 4\times 10^{-6}$, rendering any detection unlikely.} 
The detection probability, however, can be significantly boosted when {surveying all massive star clusters from the MW and neighboring galaxy clusters. Specifically, we predict $\sim 1.5-6.5$ and $\sim 40-2800$~active Pop~III XRBs in the MW and the Virgo cluster, respectively. Our Pop~III XRBs are dominated ($\sim 99\%$) by black holes with a typical mass and luminosity of $\sim 45\ \rm M_{\odot}$ and $\sim 10^{36}\ \rm erg\ s^{-1}$. }
Deep surveys of nearby {($\lesssim 30-300\ \rm Mpc$)} galaxy clusters {for such Pop~III XRBs} are well within reach of next-generation X-ray telescopes, such as ATHENA and LYNX. 

\end{abstract}




\section{Introduction} \label{sec:intro}
Heralding the end of the cosmic dark ages, the formation of the first stars, the so-called Population~III (Pop~III), signaled the beginning of complex structure formation in the Universe \citep{Bromm_2004}. These first stars formed out of the primordial H and He synthesized in the Big Bang \citep{Haiman_1996,Liu_2018}, within dark matter minihaloes with virial masses $\sim$ $10^6 \ \rm M_\odot$ at $z\sim 20-30$ \citep{Yoshida_2006}. As a consequence of this lack of metals, their properties are different from present-day stars, causing them to be more massive, typically $\gtrsim 10-20 \ \rm M_\odot$, implying that they end their lives in energetic supernova (SN) explosions, leaving massive remnants behind \citep{Abel_2002, Bromm_2002}. The SN explosions in turn created metal-rich environments in the early intergalactic medium (IGM) from which Population~I/II (Pop~I/II) stars were able to form at later times \citep{Ostriker_1996, Heger_2002, Karlsson_2013, Jeon_2015}. 

Cosmological hydrodynamic simulations show that Pop~III star formation likely ended around $z \sim 6$ \citep{johnson_2013,Xu_2016, Sarmento_2018, Liu_2020_end}. This may be within the capabilities of next-generation telescopes, depending on the intrinsic brightness of these sources. However, even frontier telescopes that are set to launch soon, such as the \textit{James Webb Space Telescope} (JWST), will not be able to detect individual minihaloes at $z>10$ \citep{Zackrisson_2011,Zackrisson_2015}, due to their low luminosities \citep{Rydberg_2013}. Rather, a futuristic telescope such as the \textit{Ultimately Large Telescope} (ULT), a 100m liquid-mirror telescope on the Moon, would be necessary to image these exceedingly faint objects \citep{Schauer_2020}.

Given the challenges to detect the first stars directly at high redshifts, with any telescope capable of doing so, such as the ULT, lying far in the future, we must turn to indirect ways to elucidate their properties, and to test our respective theories. One approach aims at their compact, neutron star (NS) and black hole (BH), remnants, which persist throughout cosmic history, allowing us to probe them in the local Universe. 
{
For instance, the gravitational waves from the mergers of Pop~III remnants have been intensively studied (e.g. \citealt{kinugawa2021,hijikawa2021population,liu2021gravitational}), and found to be able to explain the most massive events detected by the Laser Interferometer Gravitational-Wave Observatory (LIGO) campaign \citep{abbott2020population,3ogc,lmbh2021}. 

In addition to gravitational wave sources, Pop~III compact object remnants\footnote{The Pop~III initial mass function is also predicted to reach the low mass regime of white dwarf progenitors. Here, we will not consider white dwarf remnants \citep[but see][]{Chabrier_2004}.} can also become luminous in X-rays when they experience strong tidal encounters with a companion star to form an X-ray binary (XRB) or a tidal disruption event (TDE). 
In this study, we investigate the feasibility of detecting Pop~III XRBs/TDEs from tidal captures/disruptions in our Milky Way (MW) Galaxy and nearby galaxy clusters such as Virgo. 
}

The paper is structured as follows. In Section~\ref{sec:methods}, we employ a variant of the Press-Schechter formalism, regulated by Lyman-Werner (LW) feedback, to calculate the total mass of Pop~III stars that have ever formed in the co-moving volume of the MW. Assuming a widely used prediction for the Pop~III IMF, we estimate the number of {(compact object) remnants} formed per unit stellar mass. In Section~\ref{sec:xray}, we consider {how these remnants can be detected in XRBs and distinguished from those originating in Population~I and II (Pop~I/II), based on their tidal capture and disruption rates}, as well as X-ray luminosities. We offer final conclusions in Section~\ref{sec:conclude}.

\section{Cosmological Context} \label{sec:methods}

\subsection{Halo Mass Function}
\label{sec:hmf} 

\begin{figure}
	\includegraphics[width=\columnwidth]{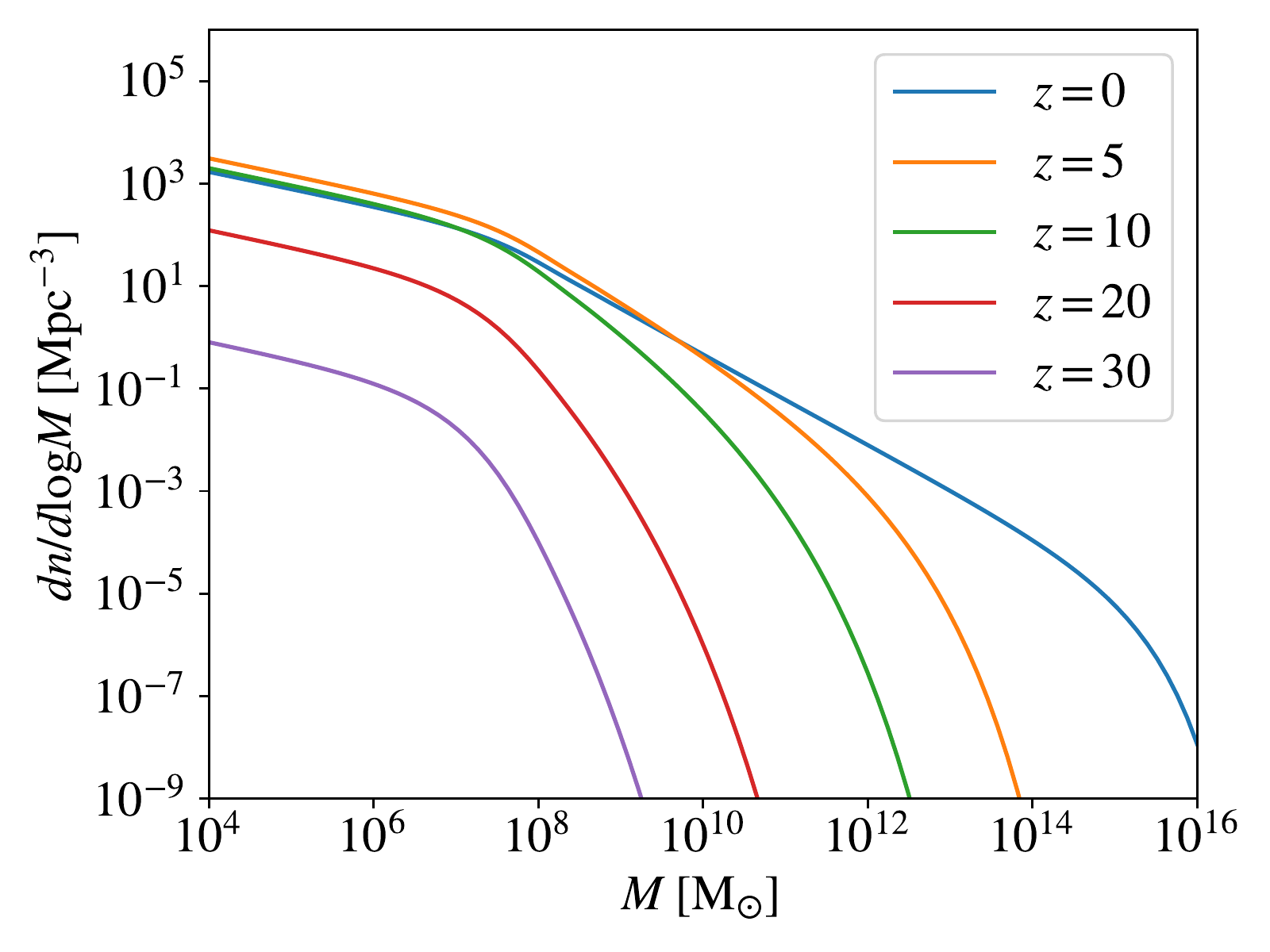}
    \vspace{-20 pt}
    \caption{Comoving number density of dark matter haloes per dex of halo mass for different redshifts. {The halo mass function is derived with the Press-Schechter formalism considering ellipsoidal dynamics, based on the \textit{Planck} $\Lambda$CDM power spectrum \citep{planck}, provided by the Python toolkit \textsc{colossus} \citep{diemer2018}}.}
    \label{fig:mhmf_figure}
\end{figure}

Within the $\Lambda$CDM model of cosmological structure formation, we can determine the number density of dark matter haloes as a function of halo mass, $M$, and redshift, $z$. A convenient way to capture the fundamental bottom-up nature of $\Lambda$CDM is provided by variants of the Press-Schechter formalism, expressing the halo mass function as \citep{Press_1974}:

\begin{equation}
    n(M,z)\,\text{d}M = \frac{\overline{\rho}}{M^2}f_{\text{EC}}(\nu)\left| \frac{\text{d\,ln\,}\nu}{\text{d\,ln\,}M}\right|\text{d}M\mbox{\ .}
	\label{eq:ps}
\end{equation}
Here, $\nu$ expresses the degree of overdensity within the Gaussian field of random density fluctuations in the underlying dark matter component. We use $f_{\text{EC}}(\nu)$ as a modification to the original Press-Schechter formalism in order to account for ellipsoidal dynamics within haloes \citep{Mo}. In Figure~\ref{fig:mhmf_figure}, we show the resulting mass functions for a number of redshifts, {based on the power spectrum of density perturbations measured by the \textit{Planck} satellite \citep{planck}}.

Approximating the pre-virialization volume of the present-day MW before breaking away from the general Hubble flow as a sphere with a (comoving) radius of $\sim$\,2~Mpc, we can integrate Equ.~\ref{eq:ps} over the range of redshift and halo mass appropriate for Pop~III star-forming minihaloes. Thus, we estimate the total stellar mass of Pop~III stars that have ever formed in the MW as follows:
\begin{equation}
\begin{split}
     M_{\star,\rm PopIII}  &\simeq V_{\text{MW}}f_{\text{b}}f_{\text{SF}}\\
     &\times \int_{5}^{30}\int_{10^6\ \rm M_{\sun}}^{10^8 \rm\ M_{\sun}} \text{max}\left[-\frac{\text{d}n(M,z)}{\text{d}z},0 \right]M\text{d}M\text{d}z \mbox{\ ,}
	\label{eq:nmh}
\end{split}
\end{equation}
where $V_{\text{MW}}$ is the approximate comoving volume of the MW ($\sim$ 30 Mpc$^3$), $f_{\text{b}}$ is the cosmic baryon mass fraction ($\sim$ 17\%), and $f_{\text{SF}}$ is the star formation efficiency ($\sim$ 1\%). The lower bound of minihalo mass ($10^6 \ \rm M_\odot$) is the critical mass for H$_2$ cooling to be efficient, considering the effect of baryon-dark matter streaming motions \citep{anna2019}. The upper bound ($10^8 \ \rm M_\odot$) is the typical halo mass for the first galaxies above which haloes are mostly metal enriched and thus do not form Pop~III stars. { This upper bound is closely connected to the $T_{\rm vir} \sim 10^4$ K atomic cooling threshold \citep[e.g.][]{Oh_2002,BrommY_2011}. The Pop III star formation rate history predicted by our simple model, based on halo mass functions (especially when taking into account LW feedback), is consistent with the results of cosmological simulations \citep[e.g.][and references therein]{Liu_2020_end}.} Here, we assume for simplicity that Pop~III star formation is terminated after reionization at $z=5$. 

Using this formalism, we find a total stellar mass of $\sim 10^7\rm\ M_{\sun}$ from $\sim$ 17,000 Pop~III forming minihaloes in the MW. As we will see below, this provides a robust upper limit on the Pop~III stellar content in the MW, which in reality is reduced by a number of feedback effects.

\subsection{Lyman-Werner Feedback}
\label{s2.2}

Specifically, we must take into account Lyman-Werner (LW) feedback in the early Universe, which leads to a reduction in Pop III forming minihaloes and stellar mass. This feedback is caused by the LW background, which consists of soft UV photons just below the hydrogen ionization threshold. These photons are energetic enough to dissociate H$_2$, thus preventing gas from cooling \citep{Oh_2002, Bromm_2003}. Hot gas is unable to form Pop~III stars, until sufficient shielding is achieved so that H$_2$ can be reformed, for LW fluxes that are not too strong \citep{Wolcott-Green_2017, Regan_2018}. In turn, this would reduce the overall number of NS and BH relics that exist today. {We model this feedback from LW flux as done in \citet{Greif_2006}:
\begin{equation}
    f_{\rm{LW}}(M, J_{21}(z))= 
    \begin{cases}
    0,& \text{if } M < M_{\rm{crit}}\\ 
    0.06\, {\rm{ln}} (M/M_{\rm crit}),& \text{if } M >M_{\rm{crit}}\ ,
    \end{cases}
\end{equation}
where $M$ is in units of M$_{\sun}$, and the critical mass is given by}
\begin{displaymath}
M_{\rm{crit}}=\left[1.25 \times 10^5 + 8.7 \times 10^5 F_{21}(z)\right]\ \rm M_{\odot} \mbox{\ .}
\end{displaymath}
Here, $F_{21}(z) = 4\pi J_{21}(z) $ is the flux in the LW bands, related to the intensity $J_{\rm LW}$, which is expressed in the standard units of $J_{\rm LW}=J_{21}\times10^{-21}\ \rm erg\ s^{-1}\ Hz^{-1}\ cm^{-2}\ sr^{-1}$, taken from \citet{Liu_2020_end}. { In our case, LW feedback is most effective for haloes with $M\lesssim 10^7\, \rm M_{\sun}$ at $z\lesssim 15$, where $f_{\rm LW}\lesssim 0.1$. It is least effective for haloes approaching the atomic cooling threshold ($M\sim 10^8\, \rm M_{\sun}$), where $f_{\rm LW}\sim 0.25-0.4$ for $z\sim 5-30$.} We have verified that our results are not sensitive to variations of the LW flux normalization within an order of magnitude.

We can now integrate the Press-Schechter function, incorporating this reduction factor, to assess how LW feedback affects the total Pop~III stellar mass ever formed in the MW:
\begin{equation}
\begin{split}
     M_{\star,\rm PopIII} \simeq V_{\text{MW}}&f_{\text{b}}f_{\text{SF}}\\ \times\int_{5}^{30}\int_{10^6\rm\ M_{\sun}}^{10^8\rm\ M_{\sun}} &\text{max}\left[-\frac{\text{d}n(M,z)}{\text{d}z},0 \right]M f_{\rm{LW}}(M, J_{21})\,\text{d}M\text{d}z \mbox{\ .}
	\label{eq:nmh-lwt}
\end{split}
\end{equation}
Evaluating this expression, we find $M_{\star,\rm PopIII}\simeq 2.3\times 10^{6}\ \rm M_{\sun}$ from $\sim 10,000$ minihaloes. It is evident that the resulting reduction is significant, and it would be further enhanced if additional feedback processes, such as mechanical SN feedback, were considered for a more complete treatment \citep{Ciardi_2005,Jeon_2014}. However, on the scale of minihaloes prior to reionization, LW feedback is the most important effect in setting global number densities. 


\subsection{Pop~III Compact Remnant Formation}
\label{s2.3}

In order to estimate the total number of Pop~III remnants from a given total Pop~III stellar mass, we need to know the Pop~III IMF and the final fates of Pop~III stars with different initial masses. 
Numerical simulations have shown that the Pop~III IMF is top-heavy \citep[e.g.][]{Stacy2016}, with typical masses of a few 10~$\rm M_{\sun}$, but extending to significantly lower and higher masses, as well. We here assume a minimum mass of $m_{1}=1~\rm M_{\sun}$ and a maximum of $m_{2}=100~\rm M_{\sun}$, but our results do not depend sensitively on these limits. For simplicity, we represent the Pop~III IMF normalized by the total stellar mass with a simple power-law form:
\begin{equation} \label{eq:imf3}
    \frac{dN_{\star}}{dm_{\rm i}} = A m_{\rm i}^{-x}\mbox{\ ,}
\end{equation}
where $x = 1.35$ for a top heavy IMF, and $A=(2-x)/[m_{2}^{2-x}-m_{1}^{2-x}]$ is the normalization factor defined by $\int_{m_{1}}^{m_{2}} m_{\rm i}(dN_{\star}/dm_{\rm i})dm_{\rm i}=1$. For this IMF, the average stellar mass is $\bar{m}_{\rm i}\simeq 12.7\rm M_{\sun}$.

Generally, the final fates of stars can be summarized as follows, according to the initial stellar mass $m_{\rm i}$:
\begin{align*}
     1\ \mathrm{M_{\rm \odot}}< &m_{\rm i} < 10\, \rm M_{\sun}  \xrightarrow[]{} \text{White Dwarf}\\
    10\, \mathrm{M_{\sun}}< \,&m_{\rm i} < M_{\rm c}  \xrightarrow[]{} \text{Neutron Star}\\
     M_{\rm c} < \,&m_{\rm i} < 100\,\rm M_{\sun}  \xrightarrow[]{} \text{Black Hole}
\end{align*}
Here, $M_{\rm c}$, is the cutoff mass between NS and BH progenitors, which is poorly known for Pop~III, whereas $M_{\rm c}\simeq 35\rm M_{\sun}$ for metal-rich (Pop~I) stars \citep[e.g.][]{Burrows_2021}.

{ Here we have set the lower mass limit for NS formation to 10 M$_{\sun}$, which corresponds to $m_{\rm f}\sim 1.4\ \rm M_{\odot}$ from the stellar evolution models in \citet{Heger_2002}. However, since mass loss due to stellar winds decreases with metallicity, the properties of the progenitor star will likely be impacted, making it possible that this limit is lower at lower metallicities \citep[][]{2001A&A...369..574V, 2021MNRAS.504.2051V, 2012A&A...542A.113Y}. Actually, it has been argued that progenitor stars of $7-8$~M$_{\sun}$ can actually form NSs in the low-metallicity environments characteristic for Pop~III star formation \citep[][]{2012ApJ...749...91F}. Mechanisms like electron capture can also cause lower mass progenitors ($\sim6-8\ \rm M_{\sun}$) to form NS remnants \citep[][]{2019MNRAS.482.2234G}. However, we will show below that varying the lower mass limit from 10~M$_{\sun}$ to 6~M$_{\sun}$ does not significantly change the number of NSs formed per unit Pop~III stellar mass. }


\begin{figure}
	\includegraphics[width=\columnwidth]{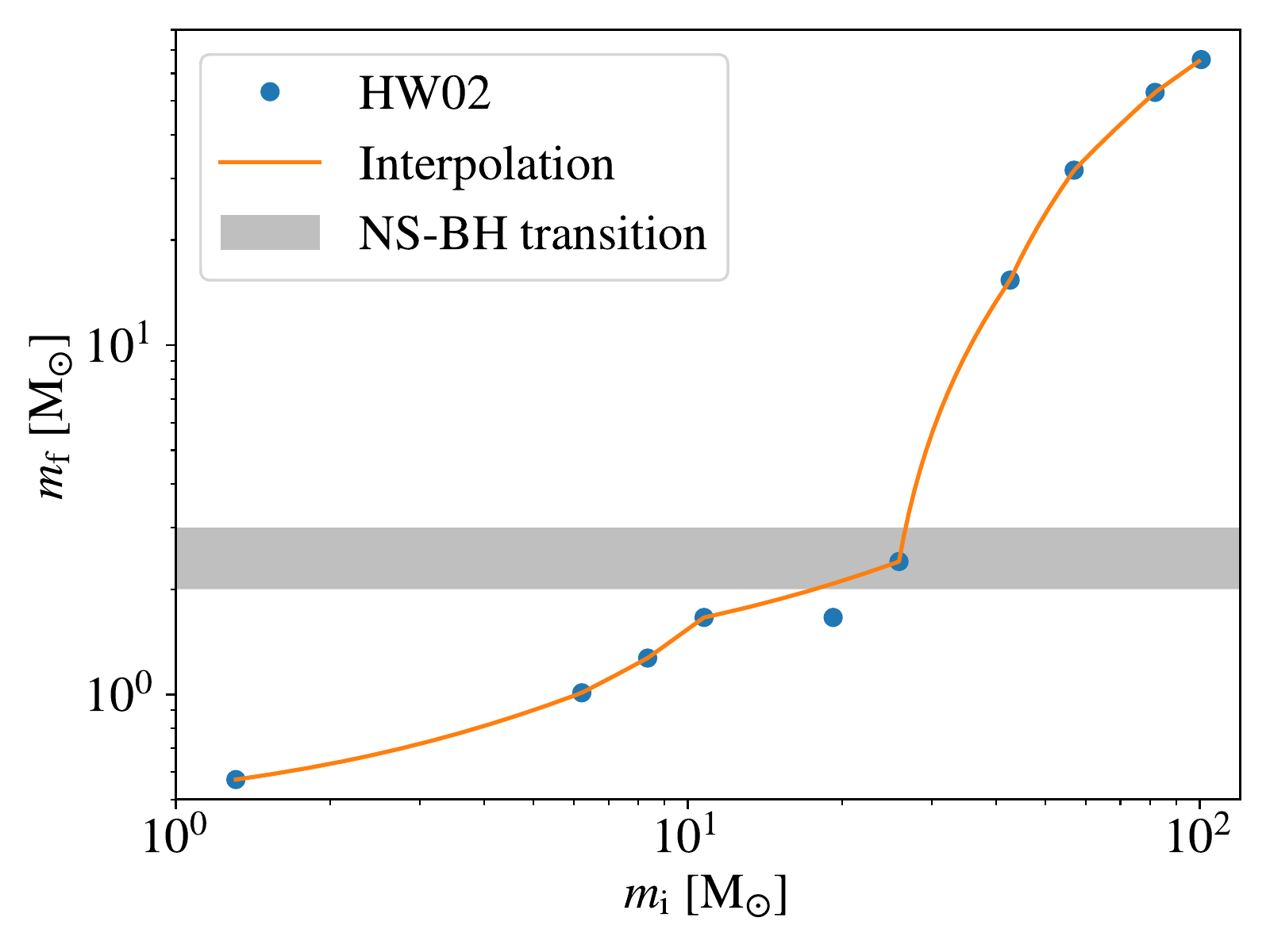}
    \vspace{-20 pt}
    \caption{ Initial-remnant mass relation from the stellar evolution models in \citet[HW02, data points]{Heger_2002} and our interpolation (solid curve). The shaded region shows the upper mass limit for NSs, reflecting its uncertain value within a range of $2-3\ \rm M_\odot$. Here we have excluded the data point at $ m_{\rm i}\simeq 19\ \rm M_{\odot}$, $m_{\rm f}\simeq 1.7\ \rm M_{\odot}$ in the interpolation to avoid divergence in $d m_{\rm i}/dm_{\rm f}$. This has little impact on our results.}
    \label{fig:mtransf}
\end{figure}

To further derive the distribution of remnant mass $m_{\rm f}$, we use the initial-remnant mass relation {for zero-metallicity stars} from~\citet{Heger_2002}, shown in Figure~\ref{fig:mtransf}. 
Considering the upper mass limit for NSs $\sim 2-3\ \rm M_{\sun}$, we find that the NS-BH transition mass is $M_{\rm c}\sim 18-27\, \rm M_{\sun}$. Note that when mapping $m_{\rm i}$ to $m_{\rm f}$ we have ignored the effect of close ($a\lesssim 10\ \rm au$) binary interactions (e.g. mass transfer and mergers). The fraction and properties of (close) Pop~III binaries are still in debate and beyond the scope of this study (see e.g. \citealt{stacy2013constraining,ryu2015formation,kinugawa2014possible,belczynski2017likelihood,liu2021binary}). Actually, recent hydrodynamic simulations of Pop III star formation and N-body simulations of young Pop III groups have found that close binaries of Pop III stars are likely rare due to the expansion of the system by angular momentum conservation during protostellar accretion \citep{heath2020,sugimura2020birth,liu2021binary,park2021}. 

Now we define the NS formation efficiency, $f_{\rm NS}$, as the number of NSs formed per unit Pop~III stellar mass. Once the cut-off mass $M_{\rm c}$ is known (from the initial-remnant mass relation), $f_{\rm NS}$ can be derived from the IMF as follows (with the normalization employed here):
\begin{equation}
    f_{\rm NS}=\int_{10\ \rm M_{\odot}}^{M_{\rm c}}\frac{dN_{\star}}{dm}dm \mbox{\ ,}
\end{equation}
giving { $f_{\rm NS} = 9.4 \times 10^{-3} \ \rm M_{\sun}^{-1}$ for $M_{\rm c} = 20 \ \rm M_\odot$.} If $M_{\rm c} = 35 \ \rm M_\odot$, we have $f_{\rm NS} = 1.5 \times 10^{-2} \ \rm M_{\sun}^{-1}$ instead. For completeness, we can also define a formation efficiency, $f_{\rm rem}$, for all compact object remnants, including both NSs and BHs (with $ m_{\rm i}>10\ \rm M_{\odot}$). Assuming our top-heavy Pop~III IMF, we find $f_{\rm rem} = 2.4 \times 10^{-2} \ \rm M_{\sun}^{-1}$, reflecting the dominance of BH remnants for Pop~III.

{ If we assume that the lower mass limit for NS formation is 6 M$_{\sun}$ (corresponding to $m_{\rm f}\sim 1\ \rm M_{\odot}$ in \citealt{Heger_2002}), we find that $f_{\rm NS} = 1.8 \times 10^{-2} \ \rm M_{\sun}^{-1}$ for $M_{\rm c} = 20 \ \rm M_\odot$, an increase by a factor of 2 from the value with a lower mass limit of 10 M$_{\sun}$. Similarly, we would have $f_{\rm rem} = 3.3 \times 10^{-2} \ \rm M_{\sun}^{-1}$, an increase by $\sim 40$\% from the 10 M$_{\sun}$ case. Actually, it is found that varying the mass limit for NS formation from $10\ \rm M_{\odot}$ to $6\ \rm M_{\odot}$ only increases the number of NS XRBs by $\sim13\%$.}

\section{Remnant X-ray Signature} \label{sec:xray}
\subsection{Distinguishing Remnants}

{ How could we distinguish whether a given observed compact object originated from Pop~I/II or Pop~III? For BHs, according to the no-hair theorem, we must rely on the difference in mass and spin of BHs from different stellar populations. One important feature in the mass spectrum of stellar-mass BHs that can be utilized is the mass gap of pair-instability supernovae (PISNe). The standard PISN model for Pop~I/II stars predicts a gap in BH mass at $\sim 55-130\ \rm M_{\odot}$ (see e.g. \citealt{belczynski2016effect, woosley2017pulsational,marchant2019pulsational}). However, the gap is narrowed to $\sim 85-120\ \rm M_{\odot}$ for Pop~III stars due to their compactness and strongly reduced mass loss \citep{farrell2020gw190521,tanikawa2020population}. Therefore, (low-spin)\footnote{Mergers of Pop~I/II BHs can also result in masses $\sim 55-85\ \rm M_{\odot}$. However, such BHs will also gain high spins from the mergers.} BHs in the mass range $\sim 55-85\ \rm M_{\odot}$ may originate mainly from Pop~III stars \citep{kinugawa2021formation,bl2020gw190521,liu2021gravitational}, as those inferred in the special event GW190521 with masses $85_{-14}^{+21}\ \rm M_{\odot}$ and $66_{-18}^{+17}\ \rm M_{\odot}$ \citep{abbott2020gw190521}. The top-heavy IMF assumed in this work predicts that $\sim 5\%$ of Pop~III remnants will have masses in this range. It will be shown below that the fraction is even higher in XRBs.

For NSs, there are additional physical properties to consider and the situation is also more complex.}
There are approximately 100 million NSs in the MW \citep{numns}, based on the number of stars having undergone SN explosions. However, most of these NSs radiate very little energy, making them extremely hard to detect. In general, the NSs that we have been able to detect are pulsars \citep{manchester2005australia}, i.e. highly magnetized, rapidly rotating NSs which emit in a beamed polar pattern, or those in binary systems. The latter can also be detected through their gravitational wave emission in rare cases {\citep[GW170817, GW200105 and GW200115;][]{Abbott_2017,abbott_2021}.} In Sec.~\ref{s2.2}, we calculated that there are {$M_{\star,\rm PopIII}\sim 2.3\times 10^{6}\ \rm M_{\odot}$} of Pop~III stars ever formed in the comoving volume of the MW, leading to {$\sim 1.6\times 10^{4}$} Pop~III NSs given a formation efficiency of $f_{\rm NS}\simeq 7\times 10^{-3}\ \rm M_{\odot}^{-1}$. 
If Pop~III NSs are equally detectable as their Pop~I/II counterparts, we would need to survey around {6,000} NSs to have a chance of finding one originating in Pop~III. In this case, location within the MW could be a promising characteristic to distinguish Pop~I/II NSs from Pop~III ones. Pop~I/II remnants tend to concentrate in the Galactic disk, while Pop~III remnants are distributed throughout the halo. This reflects the formation of Pop~III stars in dark matter minihaloes that merged to build up the MW halo, while Pop~I/II stars formed mostly in areas of the MW with dense concentrations of gas. Therefore, if we observe pulsars in the Galactic halo, we are likely to have detected Pop~III NSs. 

Unfortunately, not all NSs can be observed as pulsars, particularly for ancient Pop~III NSs. {Formed mostly at $z\gtrsim 5$, i.e. $\gtrsim 12.6\ \rm Gyr$ ago, Pop~III NSs are expected to have spun down, due to the conversion of rotational energy into magnetospheric emission \citep[e.g.][]{Ostriker_1969}. Similarly, any initially present strong magnetic fields will have decayed as a consequence of conductive dissipation.} 
The pulsar window of observation may thus not be promising for Pop~III stars, and location therefore is probably not a feasible way to distinguish these remnants. It is also important to note that {Pop~I/II origin NSs} in the MW can have large proper motions due to the asymmetric ejection of mass and neutrinos during their supernova explosions \citep{2005MNRAS.360..974H, 2017ApJ...837...84J}. Some NSs can receive birth kicks up to several hundred km~s$^{-1}$, raising the possibility that a Pop~I/II NS formed in the Galactic disk can eventually end up in the Galactic halo \citep{2018MNRAS.479.3094R}. This further eliminates location as a feasible way of distinguishing remnants as all populations of NSs exist in the Galactic halo.

{In addition, similar to BHs, there can be a statistical difference between the masses of Pop~I/II and Pop~III NSs, resulting from different IMFs and a change in the maximum NS mass due to different SN conditions. However, such differences are rather complex and can only be evaluated with a sufficiently large sample of remnants and accurate mass measurements. Just like it is difficult to observe isolated Pop~III BHs, it is also challenging to observe isolated Pop~III NSs as they are not likely to be pulsars. Nevertheless, we may be able to detect them in XRBs\footnote{To study NSs and BHs separately with XRBs, it is important to distinguish NS and BH components in XRBs, which can be achieved by observing the XRB spectra. For a NS, we expect a black-body component once the system is in thermal equilibrium due to the fact that NSs have actual surfaces with which the accreting matter collides. In contrast, BHs have no actual surfaces, but only event horizons, such that they are never able to come into thermal equilibrium \citep{Titarchuk_2005}.}. }

{ To distinguish XRBs involving Pop~III NSs with their Pop~I/II counterparts, a clearer understanding of any unique spectral signature of Pop~III XRBs is important. However, this requires sophisticated follow-up work, beyond the scope of this paper. Here we only briefly discuss the role of magnetic fields for solving this problem. As mentioned above, the magnetic fields of Pop~III NSs are expected to be weak at present. If a Pop~III NS finds itself in an XRB at some point in cosmic history after capturing of a companion star (see the following sections), the Alfv\'{e}n radius will likely be smaller than the NS radius (or the radius of the last-stable orbit) due to the lack of a strong magnetic field \citep{Lamb_1973}. As a result, we expect the accretion flow being deposited close to the equatorial stellar surface, giving rise to thermal emission, possibly modulated by thermonuclear flashes, such as those responsible for X-ray bursters \citep[e.g.][]{Joss_1984}. While for a sub-class of Pop~I/II sources with strong magnetic fields, the accreting matter can be guided by the magnetic fields to polar hot spots, which stabilize the nuclear burning shells against thermonuclear flashes \citep{Joss_1984}. Therefore, identification of Pop~III XRBs may focus particularly on bursting sources. }

{Despite the complexity in distinguishing different populations, it is important to investigate the abundance and detectability of XRBs involving Pop~III remnants in the local universe. In the following, we calculate the rate of Pop~III NSs and BHs tidally capturing or destroying a star to produce an XRB or TDE. We also derive the resulting XRB luminosity functions for the MW and the Virgo Cluster, which are then compared with current observations of XRBs.}

{
\subsection{Tidal Capture and Disruption} 
\label{sec:lum}
During a close encounter between a compact object and a star, orbital energy is transferred to internal energy of the star by tidal dissipation. The star can be destroyed leading to a luminous (but short-lived) TDE. If the star survives, an XRB can form. As such close encounters are only frequent in dense stellar environments like nuclear star clusters (NSCs), below we focus on the Galactic centre (GC), i.e. the MW NSC, as a starting point.

For simplicity, we did not model the detailed (evolution of) density and velocity structure of the MW NSC. Instead, we introduce the effective stellar density $n_{\star}$ and (1D) velocity dispersion $\sigma_{\star}$ to calculate the \textit{average} capture and disruption rates. We assume that $n_{\star}\propto m_{\rm f}^{\alpha}$ and $\sigma_{\star}\propto m_{\rm f}^{\beta}$, in which the power-law dependence on $m_{\rm f}$ is meant to capture the effect of mass segregation. To determine the normalization and power-law indexes of $n_{\star}(m_{\rm f})$ and $\sigma_{\star}(m_{\rm f})$, we calibrate our calculation to the Fiducial GC model based on the Fokker-Planck formalism in \citet{generozov2018}. 

The capture/disruption rate for a compact object of mass $m_{\rm f}$ surrounded by stars of mass $m_{\star}$ and radius $r_{\star}$ can be written as
\begin{align}
    &\Gamma'_{i}=n_{\star}\int_{0}^{v_{i}}\Sigma_{i}v f(v,\sigma_{\star})dv\ ,\quad i=\text{cap, dis}\ ,\label{gamma}\\
    &\Sigma_{\rm dis}=\pi r_{\rm dis}^{2}\left[1+\frac{2G(m_{\rm f}+m_{\star})}{v^{2}r_{\rm dis}}\right]\ ,\quad v_{\rm dis}= v_{\rm esc}=\sqrt{\frac{2Gm_{\star}}{r_{\star}}}\ ,\\
    &\Sigma_{\rm cap}=\pi r_{\rm cap}^{2}\left[1+\frac{2G(m_{\rm f}+m_{\star})}{v^{2}r_{\rm cap}}\right]-\Sigma_{\rm dis}\ ,\quad \Sigma_{\rm cap}(v_{\rm cap})=0\ .
\end{align}
Here $f(v,\sigma_{\star})$ is the relative velocity distribution, assumed to be Maxwellian with a scale parameter equal to $\sigma_{\star}$, $r_{\rm dis}=1.5 r_{\rm t}$ is the maximum pericenter radius for a star to be disrupted, given the tidal radius $r_{\rm t}=(m_{\rm f}/m_{\star})^{1/3}r_{\star}$, $r_{\rm cap}\equiv r_{\rm cap}(v,m_{\rm f}|m_{\star}, r_{\star})$ is the maximum pericenter radius for sufficient tidal energy transfer. We calculate $r_{\rm cap}$ with the method in \citet[see their equ.~20 and Appendix A and B]{generozov2018}, based on \citet{lee1986,ivanov2001}. Following \citet{generozov2018}, we adopt $m_{\star}=0.3\ \rm M_{\odot}$ and $r_{\star}=(m_{\star}/\mathrm{M_{\odot}})^{0.8}\simeq0.38\ \rm R_{\odot}$, assuming that low-mass stars dominate in NSCs. In general, $\Gamma'\equiv \Gamma'(m_{\rm f}, m_{\star}, n_{\star}, \sigma_{\star})$ is a function of $m_{\rm f}$, $m_{\star}$, $n_{\star}$ and $\sigma_{\star}$, while it becomes a monotonic function of $m_{\rm f}$ once $m_{\star}$ is fixed, and the relation between $n_{\star}$, $\sigma_{\star}$  and $m_{\rm f}$ is known. To derive this relation, we calibrate our simple effective model (Equ.~\ref{gamma}) to the Fiducial GC model in \citet{generozov2018} by reproducing the (present-day\footnote{ By calibrating with the present-day rates, we actually fix the stellar density profile to the present-day state, while the distribution of stars is more compact in the past. This leads to underestimation of the tidal capture rate by up to a factor of 2 \citep{generozov2018}.}) capture and disruption rates for Pop I/II NSs of $m_{\rm f}=1.5\ \rm M_{\odot}$ and BHs of $m_{\rm f}=10\ \rm M_{\odot}$ (see their fig.~11 and table~4)
\begin{align}
    \Gamma'_{\rm cap}(m_{\rm f}={1.5\ \rm M_{\odot}})N_{\rm NS,PopI/II}\simeq 2\times 10^{-8}\ \rm yr^{-1}\ ,\notag\\
    \Gamma'_{\rm cap}(m_{\rm f}={10\ \rm M_{\odot}})N_{\rm BH,PopI/II}\simeq 6\times 10^{-8}\ \rm yr^{-1}\ ,\notag\\
    \Gamma'_{\rm dis}(m_{\rm f}={1.5\ \rm M_{\odot}})N_{\rm NS,PopI/II}\simeq 9\times 10^{-8}\ \rm yr^{-1}\ ,\notag\\
    \Gamma'_{\rm dis}(m_{\rm f}={10\ \rm M_{\odot}})N_{\rm BH,PopI/II}\simeq 6\times 10^{-7}\ \rm yr^{-1}\ .\notag
\end{align}
Given the total numbers in their model of NSs, $N_{\rm NS,Pop I/II}=2.3\times 10^{5}$, and BHs, $N_{\rm BH,Pop I/II}=2\times 10^{4}$, we have
\begin{align}
    n_{\star}&\simeq 6\times 10^{5}\ \mathrm{pc^{-3}}\times (m_{\rm f}/10\ \rm M_{\odot})^{1.27}\ ,\\
    \sigma_{\star}&\simeq 192\ \mathrm{km\ s^{-1}}\times (m_{\rm f}/10\ \rm M_{\odot})^{0.23}\ .
\end{align}

Now we apply our model to Pop~III remnants to estimate the numbers of (active) XRBs and TDEs. We start with the simple case in which all remnants have the same mass $m_{\rm f}$. We fix the total number of remnants in the MW halo to $N_{\rm PopIII,halo}=f_{\rm rem}M_{\star,\rm PopIII}\sim 5.4
\times 10^{4}-2.4\times 10^{5}$, given the overall compact object formation efficiency $f_{\rm rem}=2.4\times 10^{-2}\ \rm M_{\odot}^{-1}$ (for $m_{i}>10\ \rm M_{\odot}$), where the lower (upper) bound corresponds to the case with (without) LW feedback. 
By changing $m_{\rm f}$ in this setup, we are able to demonstrate how the efficiencies of producing XRBs and TDEs depend on $m_{\rm f}$. As this is an oversimplification of the reality with remnants of different masses, we label the quantities derived in this way as effective numbers. In the next subsection, we generalize this simple analysis to consider the mass spectrum of Pop~III remnants.


Given $N_{\rm PopIII,halo}$ and $m_{\rm f}$, we first derive the total number of Pop~III remnants that fall into the GC, $N_{\rm PopIII,GC}$. According to \citet{Madau_2001}, remnants within $r_{\rm df}\sim 11\ \mathrm{pc}\times(m_{\rm f}/45\ \mathrm{M_{\odot}})^{1/2}$ will spiral into the GC by dynamical friction. Assuming that the distribution of Pop~III remnants is an isothermal sphere\footnote{ This assumption is generally consistent with cosmological simulations (see fig.~4 in \citealt{liu2021gravitational}). }, we have $N_{\rm PopIII,GC}=N_{\rm PopIII,halo}(r_{\rm df}/r_{\rm halo})\sim 6-26\times (m_{\rm f}/45\ \mathrm{M_{\odot}})^{1/2}$, given the MW halo radius $r_{\rm halo}\sim 100\ \rm kpc$. Then we write the effective numbers of active Pop~III XRBs and TDEs as
\begin{align}
    N_{\rm XRB}&=N_{\rm PopIII,GC}\Gamma'_{\rm cap}t_{\rm XRB}\ ,\label{nxrb}\\
    N_{\rm TDE}&=N_{\rm PopIII,GC}\Gamma'_{\rm dis}t_{\rm TDE}\ ,
\end{align}
where we employ the XRB lifetime $t_{\rm XRB}$ (after Roche-lobe contact) from \citet[see their fig.~10]{generozov2018}, and define the TDE lifetime $t_{\rm TDE}\simeq 63t_{\rm Edd}$ as the time when the luminosity has dropped to $10^{-3}$ of the Eddington luminosity. Here, $t_{\rm Edd}$ is the timescale of super-Eddington accretion, calculated with the method in \citet[see their equ.~4]{piran2012}. We have $t_{\rm TDE}\sim 700-3400\ \rm yr$ for $m_{\rm f}\sim 1.4-65\ \rm M_{\odot}$. 
\begin{figure}
    \centering
    \includegraphics[width=\columnwidth]{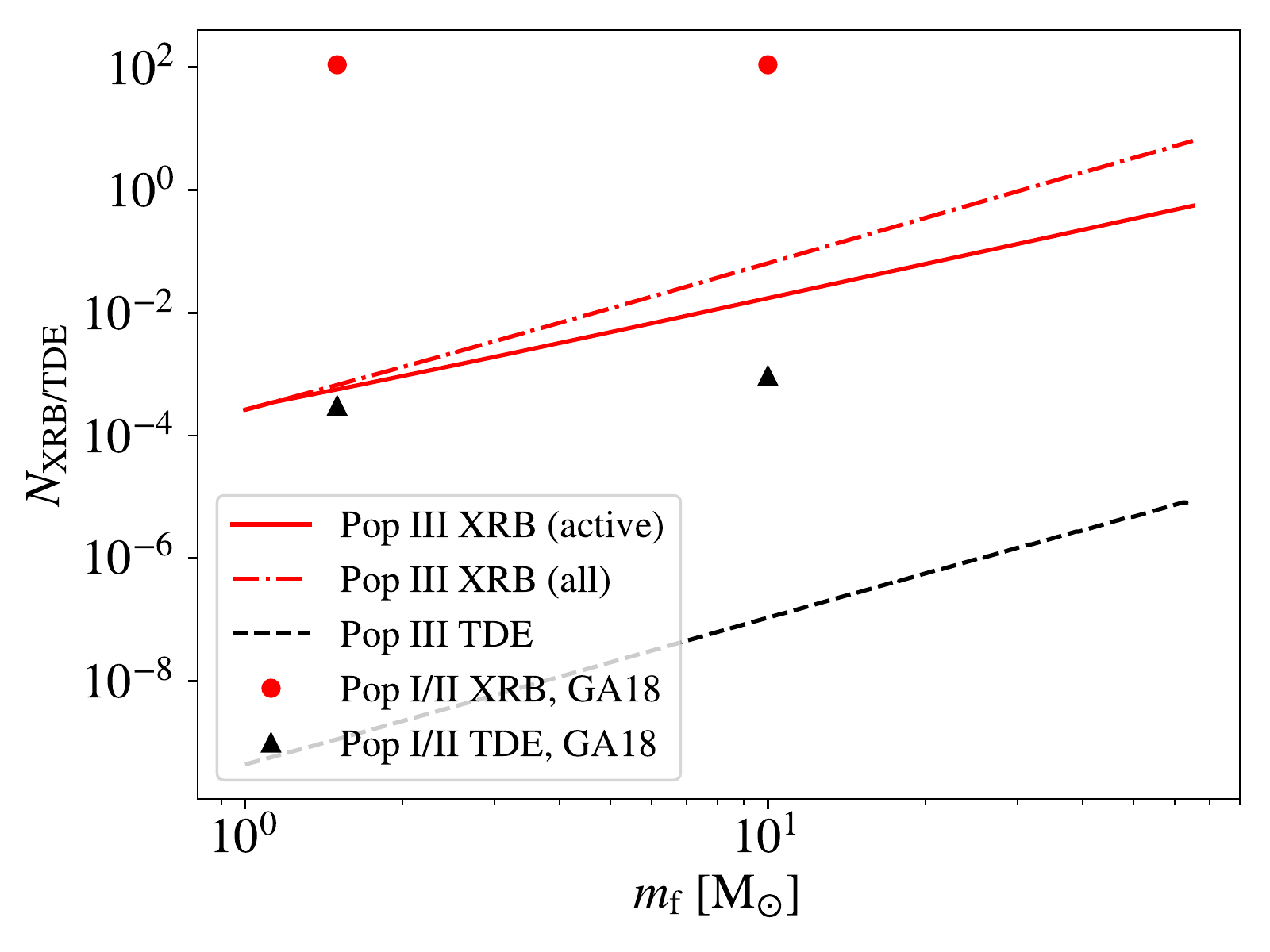}
    \vspace{-20 pt}
    \caption{ Effective numbers of Pop~III XRBs (solid curve) and TDEs (dashed curve) in the GC with LW feedback. The effective number of all Pop~III XRBs ever formed in the GC is shown with the dashed-dotted curve. 
    For comparison, we also plot the numbers of all Pop I/II XRBs (filled circles) and active TDEs (triangles) from the Fiducial model in \citet[GA18]{generozov2018}. The numbers of Pop I/II XRBs are given by table~3 of GA18, while the numbers of Pop I/II TDEs are calculated with the disruption rates in table~4 of GA18, and TDE lifetimes based on \citet{piran2012}. }
    \label{fig:ncap_m}
\end{figure}

In Figure~\ref{fig:ncap_m}, we show our results under LW feedback, in comparison with the predictions for Pop~I/II XRBs and TDEs from the Fiducial model in \citet{generozov2018}. Neglecting LW feedback would increase the numbers by a factor of 4.4. We also plot the effective number of all Pop~III XRBs ever formed in the GC by replacing $t_{\rm XRB}$ in Equ.~\ref{nxrb} with the age of the MW NSC, $t_{\rm GC}\sim 10\ \rm Gyr$. 
We find that $N_{\rm XRB}$ rises rapidly from $\sim 5\times 10^{-4}$ to $\sim 0.55$ when $m_{\rm f}$ increases from $1.4\ \rm M_{\odot}$ to $65\ \rm M_{\odot}$, showing that BHs are much more efficient than NSs in producing XRBs. The reason is that BHs fall into the GC more rapidly and also capture stars more easily in the GC due to their higher masses. Further taking into account the remnant mass spectrum,
mapping $m_{\rm i}$ to $m_{\rm f}$ with the initial-remnant mass relation used here (see Fig.~\ref{fig:mtransf}), 
we can average $N_{\rm XRB}$ over the IMF to estimate the overall number of active Pop~III XRBs as 
\begin{displaymath}
\hat{N}_{\rm XRB}=
\frac{\int_{10\ \rm M_{\odot}}^{100\ \rm M_{\odot}} (dN_{\star}/d m_{\rm i}) N_{\rm XRB}(m_{\rm f}( m_{\rm i}))d m_{\rm i}}{\int_{10\ \rm M_{\odot}}^{100\ \rm M_{\odot}} (d N_{\star}/d m_{\rm i})d m_{\rm i}}\simeq 0.06\ (0.3)\mbox{\ ,}
\end{displaymath}
with (without) LW feedback\footnote{Similarly, we can calculate the overall number $\hat{N}_{\rm PopIII,GC}$ of Pop~III remnants in the GC considering the mass spectrum by taking an IMF average of $N_{\rm PopIII,GC}(m_{\rm f}(m_{\rm i}))$. We find that $\hat{N}_{\rm PopIII,GC}\sim 3\ (11)$ with (without) LW feedback, close to the prediction $\hat{N}_{\rm PopIII,GC}\sim 4$ for remnants of $m_{\rm f}\lesssim 65\ \rm M_{\odot}$ from the MW merger trees (considering LW feedback) in \citet{liu2021gravitational}.}.
Evidently, $\hat{N}_{\rm XRB}$ is clearly below 1. The number of all Pop~III XRBs ever formed is higher than the active number by a factor of 10, reaching $\sim 0.6-2.5$, still much lower than the number of all Pop~I/II XRBs, $\sim 220$ predicted by \citet{generozov2018}. This indicates that most Pop~III XRBs in the GC involve BHs and are inactive at present. Similar to $N_{\rm XRB}$, $N_{\rm TDE}$ increases with $m_{\rm f}$. Although $\Gamma'_{\rm dis}$ is higher than $\Gamma'_{\rm cap}$ by a factor of $\sim 4-20$, the TDE number is much lower due to the short lifetimes of TDEs. Actually, $N_{\rm TDE}\lesssim 10^{-5}\ (10^{-3})$ for Pop~III (I/II) remnants, indicating that detection is very unlikely. 

At last, we would like to point out that we again ignore the effect of binaries in counting the number of Pop~III remnants involved in tidal encounters. The effect is expected to be very small for the following reason. 
N-body simulations of Pop~III star clusters \citep{liu2021binary} find that Pop~III binaries are mostly wide systems with typical (initial) separations $a\sim 1000\ \rm au$, and the fraction of binaries increases with mass, from $\sim 0.1-1$\% at $m_{\rm i}\sim 10\ \rm M_{\odot}$ to $\sim 70$\% at $m_{\rm i}\sim 100\ \rm M_{\odot}$ (see their fig.~14). 
However, when such wide binary stars evolve into compact objects and fall into the GC, most of them will be destroyed by 3-body encounters due to the high velocity dispersion $\sigma_{\star}\sim 100-200\ \rm km\ s^{-1}$, particularly the less massive ones. 

}

\subsection{Luminosity Function of XRBs}
Now we generalize the formalism in the previous subsection to derive the luminosity function of XRBs with Pop~III remnants for not only the current MW NSC, but also all NSCs in the MW co-moving volume and the Virgo cluster. 
We start with the calculation for the current MW NSC that serves as a reference point. The luminosity function can be written as
\begin{equation}
    \frac{dN_{\rm XRB}}{dL} = \frac{dN_{\rm{XRB}}}{dm_{\rm f}} \frac{dm_{\rm f}}{dL} \mbox{\ ,}\label{eq:dldm}
\end{equation}
{where the mass function of (active) XRBs is given by
\begin{align}
    \frac{dN_{\rm XRB}}{dm_{\rm f}}=\frac{d N_{\rm PopIII,\rm GC}}{dm_{\rm f}}\Gamma'_{\rm cap}t_{\rm XRB}\ ,\label{e15}
\end{align}
and $dm_{\rm f}/dL$ can be calculated once the relation between $L$ and $m_{\rm f}$ is known. Assuming a \textit{constant} accretion rate, we have
\begin{align}
    L=\frac{\eta(m_{\star}-m_{\min})c^{2}}{t_{\rm XRB}(m_{\rm f}, m_{\star})}\ ,\quad m_{\star}=0.3\ \rm M_{\odot}\ ,\label{e16}
\end{align}
where $m_{\min}=0.1\ \rm M_{\odot}$ is the minimum companion mass below which the accretion rate drops significantly due to the change of the companion's equation of state \citep{generozov2018}, $\eta\sim 0.1$ the radiative efficiency\footnote{$\eta\sim 0.1$ is a good approximation for non-rotating BHs. However, for NS XRBs with $L\lesssim 10^{36}\ \rm erg\ s^{-1}$, the radiative efficiency can span a large range $\eta\sim 0.002-0.2$ \citep{qiao2021}. For simplicity, we assume $\eta\sim 0.1$ for both BH and NS XRBs to estimate their `typical' luminosities. }, and $c$ the speed of light. The accretion rate implied by the XRB lifetime adopted here is highly sub-Eddington, such that $L/L_{\rm Edd}\sim 1.6-7.5\times 10^{-4}$ for $m_{\rm f}\sim 1.4-65\ \rm M_{\odot}$. 

In the expression for the XRB mass function (Equ.~\ref{e15}), we have generalized the number of remnants in the GC, $N_{\rm PopIII,GC}$, to their mass function
\begin{align}
    \frac{d N_{\rm PopIII,\rm GC}}{dm_{\rm f}}=\frac{d N_{\star}}{dm_{\rm i}}\frac{d m_{\rm i}}{dm_{\rm f}}M_{\star,\rm PopIII}(r_{\rm df}/r_{\rm halo})\ ,
\end{align}
where $dN_{\star}/dm_{\rm i}$ is the IMF normalized by the total stellar mass, $dm_{\rm i}/dm_{\rm f}$ is derived from the initial-remnant mass relation (see Fig.~\ref{fig:mtransf}), and $r_{\rm df}/r_{\rm halo}\sim 10^{-4}\ (m_{\rm f}/45\ \rm M_{\odot})^{1/2}$, given the dynamical friction radius $r_{\rm df}\sim11\ \mathrm{pc}\times(m_{\rm f}/45\ \mathrm{M_{\odot}})^{1/2}$ \citep{Madau_2001} and MW halo radius $r_{\rm halo}\sim 100\ \rm kpc$. 
}

\begin{figure*}
	\includegraphics[width=\linewidth]{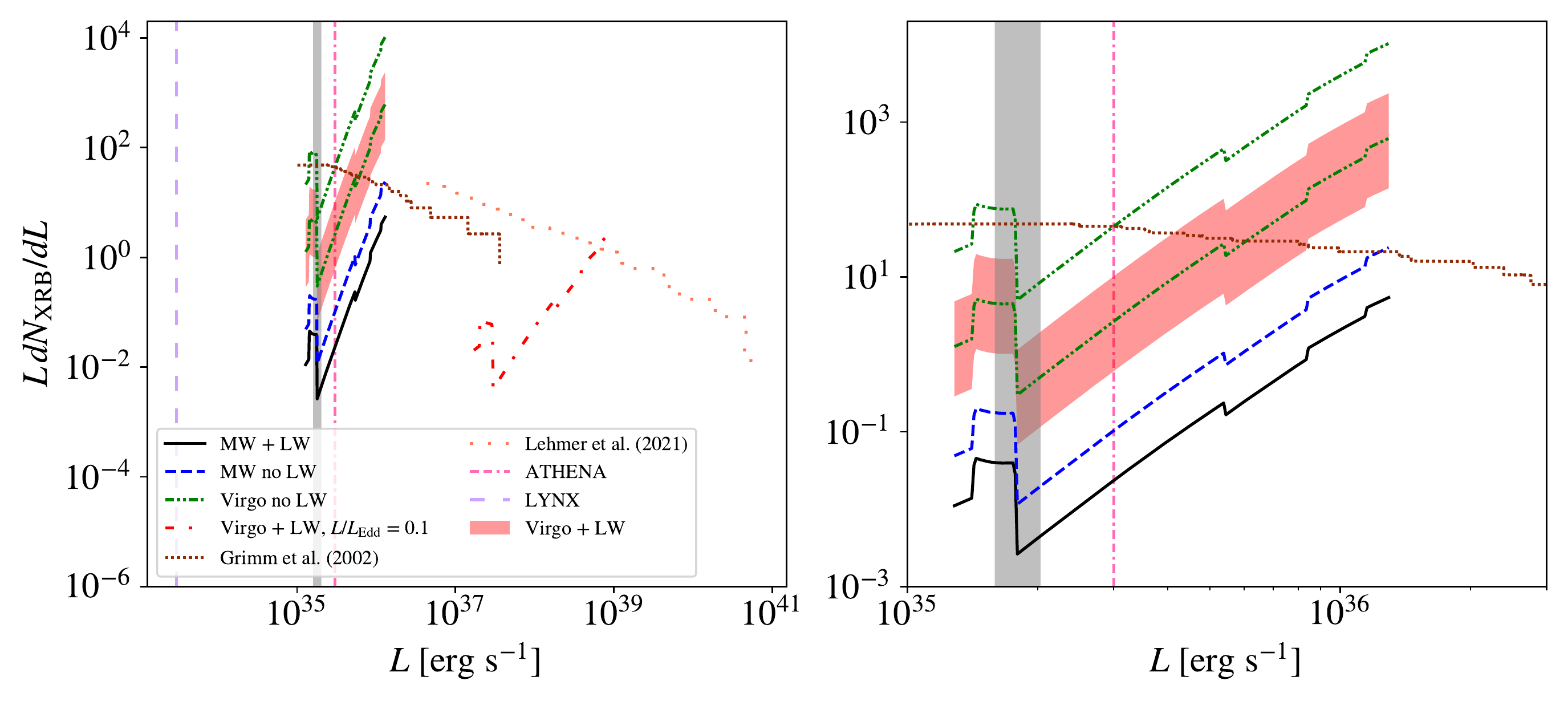}
	\vspace{-20 pt}
    \caption{Luminosity function of Pop~III XRBs. {\it Left panel:} Theoretical predictions compared with the observational data from \citet[densely-dotted]{Grimm_2002} and \citet[loosely-dotted]{Lehmer_2021} for the MW, and sensitivity limits for ATHENA (dash-dash-dotted) and LYNX (loosely-dashed) telescopes at the distance of the Virgo cluster. The results for the MW with (without) LW feedback included are shown with the solid (dashed) curve. For the Virgo cluster, the red shaded region and dash-dot-dotted curves show the results with and without LW feedback, respectively, where the upper limits come from using $\epsilon_{\rm past} = \epsilon_{\rm MW}$, while the lower limits correspond to $\epsilon_{\rm past} = 1$. The shaded grey region denotes the luminosity range corresponding to the upper mass limit for NSs. 
    For completeness, we also show the high-accretion case of $L/L_{\rm Edd}=0.1$  in the Virgo Cluster with LW feedback (dash-dotted curve). {\it Right panel:} Zoom in to the left panel in the luminosity range $L\sim 10^{35}- 3\times 10^{36}$ erg s$^{-1}$. It is evident that our predictions for the MW are below the (extrapolation of) XRB luminosity functions in observations from \citet{Grimm_2002} and \citet{Lehmer_2021}. Interestingly, without LW feedback, our Pop~III XRBs can make up a significant fraction of observed sources at $L\sim 10^{36}$ erg s$^{-1}$.}
    \label{fig:mplot}
\end{figure*}

As found in the previous subsection, the predicted number of Pop~III XRBs in the current MW NSC is less than 1, which means that we would need to observe more star clusters in order to detect a Pop~III remnant. Therefore, we further take into account the NSCs of progenitor galaxies that merged earlier in cosmic history into the MW halo, increasing the sites of Pop~III XRB capture. {
The majority of these NSCs survive the assembly of the MW (as globular clusters and massive star clusters in the MW bulge and its satellites), while a tiny fraction (within $\sim 0.5-5\ \rm kpc$ of the GC) may merge into the MW NSC and carry additional Pop~III remnants to the GC\footnote{ We only consider NSCs with masses $M_{\rm NSC}>10^{5}\ \rm M_{\odot}$ that will survive for $\sim10\ \rm Gyr$ in the MW halo \citep{zwart2010ymc}. For simplicity we do not model mergers with the MW NSC, whose effect is expected to be small due to their rareness, as the dynamical friction radius $r_{\rm df}\sim11\ \mathrm{pc}\times (M_{\rm NSC}/45\ \rm M_{\odot})^{1/2}\sim 0.5-5\ \rm kpc$ for $M_{\rm NSC}\sim 10^{5}-10^{7}\ \rm M_{\odot}$ is much smaller than the MW halo size $r_{\rm halo}\sim 100\ \rm kpc$.}. We boost our GC luminosity function (Equ.~\ref{eq:dldm}) by an efficiency factor, $\epsilon_{\rm MW}$, defined as the ratio of the total number of Pop III XRBs in the NSCs from all galaxies that end up in the MW co-moving volume to the number in the current MW NSC:
\begin{equation} \label{eq:ep_mw}
    \epsilon_{\rm MW}=\frac{\int_{M_{\star,\min}}^{\infty}\Psi(M_{\star})\Gamma'_{\rm cap}(M_{\star})dM_{\star}}{\int_{M_{\star,0}}^{\infty}\Psi(M_{\star})\Gamma'_{\rm cap,MW}dM_{\star}}\simeq 22 \mbox{\ .}
\end{equation}
Here $\Psi(M_{\star})$ is the distribution of host galaxy (stellar) mass $M_{\star}$ weighted by the number of Pop~III remnants in their NSCs, predicted by the MW merger trees in \citet[see their fig.~14]{liu2021gravitational}. Massive galaxies with $M_{\star}>M_{\star,0}=10^{10}\ \rm M_{\odot}$ represent present-day MW-like galaxies, whose NSCs are assumed to be similar to the MW NSC modelled in Sec.~\ref{s2.2}. $M_{\star,\min}=10^{5}\ \rm M_{\odot} $ is the minimum galaxy mass to host NSCs, as in \citet{liu2021gravitational}. $\Psi(M_{\star})$ captures the efficiency of in-fall into NSCs for Pop~III remnants in the MW progenitor/satellite galaxies of different masses. As a conservative estimation, we adopt the $\Psi(M_{\star})$ from the the partial NSC occupation model in \citet[see their Sec.~2.3]{liu2021gravitational}, in turn based on local observations of the NSC occupation fraction as a function of $M_{\star}$ \citep{neumayer2020}. 
$\Gamma'_{\rm cap}(M_{\star})$ is the typical tidal capture rate per remnant inside the NSC of a galaxy of mass $M_{\star}$, and the MW capture rate (see Sec.~\ref{s2.2}) for reference is denoted by $\Gamma'_{\rm cap,MW}$. To make $\Gamma'_{\rm cap}$ a function of $M_{\star}$, we associate the NSC environment (characterized by $n_{\star}$ and $\sigma_{\star}$) to $M_{\star}$, with $n_{\star}m_{\star}\sim 3M_{\rm NSC}/(4\pi R_{\rm NSC}^{3})$ and $\sigma_{\star}\sim \sqrt{GM_{\rm NSC}/R_{\rm NSC}}$, where the the NSC mass $M_{\rm NSC}$ and size $R_{\rm NSC}$ are derived from $M_{\star}$ through empirical scaling relations in local observations \citep[see equ.~8-9 in \citealt{liu2021gravitational}]{neumayer2020}. Finally, we evaluate the capture rates for different $M_{\star}$ all at $m_{\rm f}=10\ \rm M_{\odot}$. As in our GC model (Sec.~\ref{s2.2}), the stellar density at this remnant mass satisfies $n_{\star}m_{\star}\sim 3M_{\rm NSC,MW}/(4\pi R_{\rm NSC,MW})^{3}$, where $M_{\rm NSC,MW}=2.5\times 10^{7}\ \rm M_{\odot}$ and $R_{\rm NSC, MW}=3.2\ \rm pc$ are the mass and size of the MW NSC \citep{schodel2018distribution}. 
 
}


To further increase the NSC sample, we consider galaxies in the Virgo Cluster {as an example for nearby galaxy clusters. We first} scale our MW results by the number of massive (`MW-like') galaxies, again defined by $M_{\star} > 10^{10}\ \rm M_{\sun}$, and $\epsilon_{\rm MW}$, the efficiency factor to take into account the contribution of low-mass galaxies that formed these `MW-like' galaxies earlier in cosmic history. Using figure~15 from \citet{2016ApJ...824...10F}, which shows the galaxy stellar mass function of the Virgo cluster, we found 6 galaxies with $M_{\star} > 10^{10}\ \rm M_{\sun}$ in the Virgo core. 
{To make our calculation more complete, we consider the contribution of low-mass ($M_{\star}<10^{10}\ \rm M_{\odot}$) galaxies \textit{currently} in the Virgo Cluster}, using another efficiency factor, {$\epsilon_{\rm Virgo}\simeq 72.5$, defined similarly to $\epsilon_{\rm MW}$, except that $\Psi(M_{\star})$ now represents the stellar mass function of Virgo galaxies at $M_{\star}\gtrsim 10^{6}\ \rm M_{\odot}$}. Combining this with the contribution of massive galaxies, we arrive at a final efficiency factor $\epsilon_{\rm total}$ for Virgo:
\begin{equation} \label{eq:ep}
    \epsilon_{\rm total} = 6\left[\epsilon_{\rm MW}+(\epsilon_{\rm Virgo}-1)\epsilon_{\rm past}\right]\ ,
\end{equation}
{where we have further introduced a parameter $\epsilon_{\rm past}$ to consider the NSCs of the progenitors of low-mass galaxies in the Virgo cluster that captured Pop III remnants in the \textit{past}}, for which we have explored the extreme upper and lower limits, $\epsilon_{\rm MW}$ and 1, respectively. {For the optimistic case with $\epsilon_{\rm past}=\epsilon_{\rm MW}\simeq 22$, we have $\epsilon_{\rm total} \simeq 9400$, while for the conservative case with $\epsilon_{\rm past}=1$, we have $\epsilon_{\rm total}\simeq 560$.} {Clearly, as $(\epsilon_{\rm Virgo}-1)\epsilon_{\rm past}/ \epsilon_{\rm MW}\sim 3-71 \gg 1$, formation of Pop~III XRBs by tidal captures in the Virgo cluster is dominated ($\sim76-99\%$) by low-mass galaxies, where the range in values is a reflection of the uncertainties in the contribution of NSCs in the \textit{progenitors} of low-mass galaxies (captured by the parameter $\epsilon_{\rm past}$ with $1<\epsilon_{\rm past}<\epsilon_{\rm MW}$).}

Using Equations~\ref{eq:dldm}-\ref{eq:ep}, we arrive at Figure~\ref{fig:mplot}, showing $L(dN_{\rm XRB}/dL)$ as a function of $L$, {where the left panel combines our predictions with the observational data from figure~12 in \citet{Grimm_2002} and figure~3 in \citet{Lehmer_2021}, as well as sensitivity limits for ATHENA and LYNX. The right panel is a zoom in of the left, focusing on the $10^{35} - 3 \times 10^{36}$ erg s$^{-1}$ luminosity range. In comparison, \citet{Grimm_2002} uses data from the All-Sky Monitor (ASM) with sensitivity $\sim 10^{-10}$ erg\, s$^{-1}$ cm$^{-2}$ in the $2-10$ keV range,
and \citet{Lehmer_2021} uses data from Chandra (sensitivity $\sim 10^{-16}$ erg\ s$^{-1}$\ cm$^{-2}$ in the $0.5-2$ keV range).} The disagreement between the \citet{Grimm_2002} and the \citet{Lehmer_2021} curves comes from the consideration of additional physical parameters on integration scaling relations by \citet{Lehmer_2021}, such as star-formation history and metallicity. 

{
Our results for the MW (scaled by $\epsilon_{\rm MW}$) are within observational constraints, 
and it is clear that the most massive Pop~III BH XRBs could make up a significant fraction of observed XRBs at $L \sim 10^{36}$ erg\,s$^{-1}$.}
{
For the luminosity function in the Virgo Cluster (scaled by $\epsilon_{\rm total}$) accounting for LW feedback, the upper limit of the shaded region with $\epsilon_{\rm past}=\epsilon_{\rm Mw}$ is well above 1 for most luminosities, including those within the NS upper mass limit range, while the lower limit of the shaded region ($\epsilon_{\rm past}=1$) is only above 1 for higher luminosities in the BH regime. This means that we can expect to find a large number of Pop~III BH XRBs and potentially a few Pop~III NS XRB in the Virgo Cluster. 
Without LW feedback, the luminosity function will be boosted by a factor of $4.4$. 
In summary, assuming that the NS-BH transition occurs at $3\ \rm M_{\odot}$, we find that in the optimistic (conservative) case with $\epsilon_{\rm total}\simeq 9400\ (560)$, there are $\sim 5-21\ (0.3-1.2)$ Pop~III NS XRBs and $\sim 630-2760\ (37-164)$ Pop~III BH XRBs in the Virgo cluster, where the lower and upper limits correspond to the cases with and without LW feedback. 

Considering the uncertainty in the accretion rate and possible bursty accretion/emission for our XRBs, we also explore a special case with a fixed higher Eddington radio $L/L_{\rm Edd}=0.1$ for the Virgo Cluster with $\epsilon_{\rm past}=\epsilon_{\rm MW}$ and LW feedback included. This regime may not be reached in most cases, but considering a broader range of luminosities is indicative of the possibilities for accretion rates and duty cyles. 
With $L/L_{\rm Edd}=0.1$, the total number of Pop~III XRBs is significantly reduced from $\sim 630$ to $\sim 1$ due to shorter lifetimes caused by the higher accretion rates, compared with our default case with $L/L_{\rm Edd}\lesssim 10^{-3}$. Interestingly, in this case we expect to find a similar number of luminous Pop~III BH XRBs at $L\sim 10^{39}\ \rm erg\ s^{-1}$ in the Virgo cluster as currently observed in the MW \citep{Lehmer_2021}. 

}
As further discussed below, Pop~III remnants in XRBs will likely be observed in the next decades with deeper and wider X-ray surveys. Since both dynamical friction and tidal capture are more efficient for more massive remnants, detection of Pop~III NS sources is more challenging than for the BH case. 

Next-generation X-ray telescopes, such as ATHENA\footnote{\url{https://www.the-athena-x-ray-observatory.eu}} and 
LYNX\footnote{\url{https://www.lynxobservatory.com}}, both set to launch in the 2030s, will be sensitive enough to detect {XRBs involving Pop~III remnants} in the Virgo Cluster. According to the ATHENA mission proposal,
the telescope will have a flux sensitivity of $\sim$ 10$^{-17}$ erg\, s$^{-1}$ cm$^{-2}$ in the $0.5-2$ keV energy band. {Given the distance to the Virgo Cluster of $\sim$ 16.5~Mpc, and assuming constant sub-Eddington accretion (Equ.~\ref{e16}), our low-mass Pop~III (NS) XRBs with a luminosity of $\sim$ 10$^{35}$\,erg\,s$^{-1}$ will have a flux of $\sim3\times 10^{-18}$ erg\, s$^{-1}$ cm$^{-2}$, implying that only higher-mass ($m_{\rm f}\gtrsim 6\ \rm M_{\odot}$) Pop~III BH XRBs in Virgo would be detectable with ATHENA. However, the luminosity can be higher if accretion is bursty, such that some Pop~III NS XRBs may also be detectable. With limited spatial resolution of ATHENA, we will only be able to resolve the central regions of galaxies ($\sim 0.5$ kpc), not dense star clusters at the Virgo distance. However, since it is unlikely that there will be multiple Pop~III sources per star cluster, we may still be able to distinguish sources. Future work can focus on the distinguishibility of XRBs between Pop~I/II and Pop~III remnants.} According to the LYNX concept study report,
the telescope will have an even better sensitivity of $\sim$ 10$^{-19}$ erg\,s$^{-1}$ cm$^{-2}$ in the $0.5-2$ keV range, well below the fluxes of Pop~III XRBs. {In general, most of our Pop~III XRBs with $L\sim 10^{36}\ \rm erg\ s^{-1}$ are detectable at $\sim 30\ (300)\ \rm Mpc$ by ATHENA (LYNX).}

\section{Summary and Conclusions}\label{sec:conclude}
{ We explore the feasibility of detecting Pop~III remnants in nearby dense star clusters where they can form luminous XRBs and TDEs via strong tidal encounters. As a starting point, we first estimate the tidal capture and disruption rates of Pop~III remnants in the GC, finding that disruption is more efficient than capture by a factor of $\sim 4-20$ in such a high velocity dispersion environment. 
However, due to the short ($\sim 10^{3}\ \rm yr$) lifetimes of TDEs, the signals of Pop~III remnants in TDEs are very unlikely to be detected. 

We therefore focus on XRBs and further derive the XRB luminosity function, $dN_{\rm XRB}/dL$, 
in the MW and the Virgo Cluster.} At all masses/luminosities, we find that our prediction is consistent with current observations, in terms of not overproducing sources. To date, we have not been able to identify Pop~III remnants in existing surveys, likely due to the fact that they are extremely {rare ($\sim 1.5-6.5$) and significantly outnumbered by Pop~I/II sources in the MW. 
Meanwhile, we find that for the Virgo cluster, $LdN_{\rm{XRB}}/dL$ is well above 1 for the low- and high-luminosity ends in our optimistic case, leading to $\sim5-20$ NS XRBs and $\sim 600-2800$ BH XRBs. Importantly, we can conclude that there are many more BHs (by a factor of $\sim 100$) than NSs in XRBs for Pop~III, due to the top-heavy IMF and the fact that dynamical friction and tidal capture are more efficient for more massive remnants. About 40\% of our Pop~III BH XRBs contain BHs of $\sim 55-65\ \rm M_{\odot}$ in the standard PISN mass gap for Pop~I/II stars. If accurate mass measurements for compact objects in XRBs are available, we may be able to identify a sub-group of XRBs of Pop~III origin. With next-generation X-ray telescopes (e.g. ATHENA and LYNX), it is possible to detect the signature of high-mass Pop~III XRBs in nearby ($\lesssim 30-300\ \rm Mpc$) massive ($\sim 10^{15}\ \rm M_{\odot}$) galaxy clusters like Virgo. }

In predicting the Pop~III XRB luminosity function, there are a number of caveats in the calculation that could significantly affect our results. Among them are {idealized modelling of stellar environments in NSCs for tidal capture as well as accretion rate and luminosity for XRBs}, underestimating Pop~III mass loss, uncertainties in the Pop~III IMF, and modifications to the cosmological model of structure formation, which would impact the number density of the small-scale dark matter minihalo hosts for Pop~III. Moreover, there are uncertainties in observations, with possible under-counting of luminous sources \citep{jonker2021observed}. Our study therefore provides a preliminary exploration of the plausible parameter space, and needs to be followed up by more detailed work in the future.


An overall lesson from studies of early structure formation, confirmed by our work here, is that the signatures of the first stars are very much hidden behind the cumulative legacy of subsequent star formation. This applies to stellar archaeology, where surveys of metal-poor stars aim to detect the nucleosynthetic imprint of Pop~III SNe \citep[e.g.][]{Beers_2005}. Finding such chemical fossils is challenging, given how quickly any distinct Pop~III abundance yields are diluted by subsequent enrichment events \citep{Ji_2015}. Other signatures, such as the luminosity function of high-redshift galaxies, as well as their spectral energy distribution, are dominated by Pop~I/II soon after the onset of cosmic star formation \citep[e.g.][]{sarmento2018,Jaacks_2019,Liu_2020_end}. A similar challenge pertains to the search for Pop~III transients, such as pair-instability SNe and gamma-ray bursts, which are predicted to be very luminous, but rare \citep[e.g][]{Hummel_2012,magg2016}. 

One possible exception are gravitational wave merger events, involving massive Pop~III BH remnants, which can constitute up to a few percent of all compact object mergers \citep{liu2021gravitational}. Regarding any hint for Pop~III NSs,
the recent GW190814 gravitational wave detection implicates a 23 $M_{\sun}$ BH, merging with a 2.6 $M_{\sun}$ compact object \citep{ligothing}. Whether this object is an extremely high-mass NS, or an extremely low-mass BH is unclear. However, if it were a high-mass NS, this would be a possible candidate for a Pop~III NS. 

Actually, the NS-BH mass transition is still not determined with high precision. 
The upper limit of NS mass is a key parameter to derive the poorly understood NS equation of state (EoS), which is crucial to determining both macroscopic and microscopic properties of NSs, such as their internal phases and tidal deformability \citep{Oezel_2016}. The EoS reflects the internal composition of NSs, which remains an open question. For instance, current theories describe the NS core with quark-gluon plasma, hyperons (containing strange quarks) and Bose-Einstein condensate \citep{2020Natur.579...20M}. These scenarios provide different amounts of pressure support against the gravity of the star, manifested in different mass to radii ratios. Typically, the upper limit on the NS mass is assumed to be $2\rm\ M_{\sun}$ when trying to constrain the EoS. However, this may be an underestimation of the true upper limit. 

As Pop~III stars are more massive and compact than present-day stars, they are more likely to produce NSs close to the upper mass limit. Therefore, Pop~III NS remnants, if detected with a large enough sample, will be a promising probe of the true NS upper mass limit and EoS. In the present work, we predict that {$0.3-20$ Pop~III NS XRBs exist in a galaxy cluster of $\sim 10^{15}\ \rm M_{\odot}$, with a typical luminosity $\sim 10^{35}\ \rm\ erg\ s^{-1}$, assuming constant accretion of radiative efficiency $\eta\sim 0.1$. The number will be reduced if these Pop~III XRBs are X-ray bursters.} 
Such Pop~III NS XRBs are detectable with next-generation X-ray telescopes at luminosity distances $\sim 10-100\ \rm Mpc$. Given such a large observable volume that includes $\gtrsim 10^{7}$ galaxies and many massive galaxy clusters (e.g. the Coma cluster), future X-ray surveys may be able to build up a large sample of Pop~III NS XRBs, providing more accurate constraints for both Pop~III stars and the NS EoS in general. {However, this requires precise measurements of NS masses in XRBs, which is challenging. The X-ray luminosity alone is not a good indicator of mass, as observations have found ultra-luminous XRB sources containing NSs with super-Eddington accretion \citep[e.g.][]{2014Natur.514..202B, 2017ApJ...836..113P, 2021MNRAS.504..701B}. This also contributes to the difficulty/complexity in distinguishing Pop~III from Pop~I/II NS XRBs. 

Beside XRBs, observations of gravitational waves from NS-NS and BH-NS mergers can also provide constraints on the NS upper mass limit and EoS (e.g. \citealt{annala2018gravitational,landry2020nonparametric,al2021combining}). Actually, the recent study by \citet{liu2021gravitational} finds that Pop~III NSs can be detected in Pop~III BH-NS mergers (which make up $\sim 7\%$ of all Pop~III mergers) with a local merger rate density up to $\sim 0.04\ \rm yr^{-1}\ Gpc^{-3}$, and a detection rate $\sim 10-180\ \rm yr^{-1}$ for 3rd-generation detectors that can reach $z\sim10$. 
}

Overall, the `archaeological' approach to studying the high-redshift Universe, where we probe local fossils in great detail, is rapidly opening up an exciting new window into early cosmic history. Over the next decade or so, this novel window is likely to further grow in importance, with a suite of next-generation observational facilities becoming available.

{
\section*{Acknowledgements}
The authors would like to thank the referee for insightful and detailed comments.
}

\section*{Data availability}
The data underlying this article will be shared on reasonable request to the corresponding authors.






\bibliographystyle{mnras}
\bibliography{example} 





\bsp	
\label{lastpage}
\end{document}